\documentclass[preprintnumbers,amsmath,12pt,amssymb,floatfix,superscriptaddress,nofootinbib]{revtex4}
\usepackage{graphicx}
\usepackage{epsfig}
\usepackage{bm}
\usepackage{amsfonts}
\def\be {\begin{equation}}
\def\ee {\end{equation}}
\def\ba {\begin{eqnarray}}
\def\ea {\end{eqnarray}}

\begin{document}

\title{ Power Law and Logarithmic Ricci Dark Energy Models in Ho\v{r}ava-Lifshitz Cosmology}

\author{Antonio Pasqua}
\affiliation{Department of Physics,
University of Trieste, Via Valerio, 2 34127 Trieste, Italy.}
 \author{Surajit Chattopadhyay}
\affiliation{Pailan College of Management and Technology, Bengal Pailan Park, Kolkata-700 104, India.}
\author{Martiros Khurshudyan}
\affiliation{Department of Theoretical Physics,
Yerevan State University, 1 A. Manookian, Armenia.}
\author{Ratbay Myrzakulov} \affiliation{Eurasian International Center for Theoretical Physics and  Department of General $\&$  Theoretical Physics, Eurasian National University, Astana 010008, Kazakhstan}
\author{Margarit Hakobyan} \affiliation{A.I. Alikhanyan National Science Laboratory, Alikhanian Brothers St. and
Department of Nuclear Physics, Yerevan State University, Yerevan, Armenia}
\author{Artashes Movsisyan} \affiliation{R$\&$D Center of Semiconductor Devices $\&$ Nanotechnologies, Yerevan State University}

\email{toto.pasqua@gmail.com}
\email{surajcha@iucaa.ernet.in}
\email{khurshudyan@yandex.ru}
\email{rmyrzakulov@gmail.com; rmyrzakulov@csufresno.edu}
\email{margarit@yerphi.am}
\email{artmovsissyan@yandex.ru}

\begin{abstract}
In this work, we studied the Power Law and the Logarithmic Entropy Corrected versions of the Ricci Dark Energy (RDE) model in a spatially non-flat universe and in the framework of Ho\v{r}ava-Lifshitz cosmology. For the two cases containing non-interacting and interacting  RDE and Dark Matter (DM), we obtained the exact differential equation that determines the evolutionary form of the RDE energy density parameter. Moreover, we obtained the expressions of the deceleration parameter $q$ and, using a parametrization of the equation of state (EoS) parameter $\omega_D$ as $\omega_D\left(z\right)=\omega_0+\omega_1 z$, we derived the expressions of both $\omega_0$ and $\omega_1$. We interestingly found that the expression of $\omega_0$ is the  same for both non-interacting and interacting case. The expression of $\omega_1$ for the interacting case has strong dependence from the interacting parameter $b^2$. The parameters derived in this work are done in small redshift approximation and for low redshift expansion of the EoS parameter.
\end{abstract}

\maketitle

\section{Introduction}
Recent astrophysical and cosmological observations clearly indicate that the present day universe is experiencing a phase of accelerated expansion \cite{1a,1b,1c,1d,1g,1h,1-1-1-1,cmb1,cmb3,planck,sds1,sds4,xray}, which represents the second accelerated expansion experienced by our Universe after the one occurred during the inflation.
In order to find a reasonable model which is able to explain the present universe, scientists began to investigate a possible acceptable explanation of this late time accelerated expansion. Three main classes of models have been proposed till now in order to explain this observed accelerated expansion of the universe:
\begin{enumerate}
\item A cosmological constant $\Lambda$;
\item Dark Energy (DE) models;
\item  Modified Gravity theories.
\end{enumerate}
The cosmological constant $\Lambda$, which has equation of state parameter $\omega = -1$, represents the earliest and simplest theoretical candidate proposed in order to explain the accelerated expansion of the universe. It is well-known, anyway,  that $\Lambda$ suffers from two main difficulties: the fine-tuning and the cosmic coincidence problems \cite{copeland-2006}. According to the first, the vacuum energy density is about $10^{123}$ times smaller than what we observe. Instead, according to the cosmic coincidence problem, the vacuum energy and DM are nearly equal today although they have evolved independently and from different mass scales (which is a particular coincidence if no internal connections between them exist). Till nowadays, different attempts have been done in order to find a possible reasonable solution for the coincidence problem \cite{delcampo,delcampoa,delcampob,delcampod,delcampoe,delcampof}. \\
The second class of models proposed in order to give an explanation to the cosmic acceleration involves DE.  The evidence of the cosmic acceleration implies that, if Einstein's theory of General Relativity must be considered valid on cosmological scales, then the universe must be dominated by a mysterious and unknown kind of missing component with some peculiar characteristics, for example it must not be clustered on large length scales and its pressure $p$ must be negative enough in order to drive the accelerated expansion the universe is undergoing. In relativistic cosmology, the cosmic acceleration can be described with the help of a perfect fluid which pressure $p$ and energy density $\rho$ satisfy the condition $\rho + 3p < 0$. This kind of fluid with negative pressure is named Dark Energy (DE). The condition  $\rho + 3p < 0$ implies that the EoS parameter $\omega$ must obey the constrain $\omega <-1/3$, while from an observational point of view it is a difficult task to constrain its precise value. Since the fundamental theory of physics which can explain the microscopic physics of DE is unknown up to now, phenomenologists try to suggest and reconstruct different models based on its macroscopic behavior.\\
If the idea of presence of Dark components is the right one, we must have that the largest part of the total energy density of the present universe is contained in the two Dark sectors which contribute to the composition of the universe, i.e. the DE and the Dark Matter (DM), which represent, respectively, about the 70$\%$ and about the 25$\%$ of the total energy density $\rho_{tot}$ of the present day universe \cite{twothirds}. The Baryonic Matter we are able to observe with our scientific instruments as stars and atoms contributes for about the 5$\%$ of the total energy density composing the universe, while the contribution given by the radiation term can be considered practically negligible since it is a small fraction of percent of the total energy density of the universe.\\
DE models have been studied taking into account several different candidates, including quintessence, Chaplygin gas, k-essence, phantom, a time-variable cosmological constant, tachyon and a unified model of quintessence and phantom known with the name of quintom \cite{odi2,pad,A3,A144,A150,A154,A10,A12,A13,A15,A16,A164,A169,A174,A27,A34,A38,A204,A205,A211,A216,A217,A219,A49,A53,A56,A57,A59,A224,A60,A226,A225,A61} even if in order to have a complete detailed description of the DE nature we need a deeper comprehension of the quantum gravity theory, which is not yet available.\\
In scientific literature, the Holographic DE (HDE) model (based on the holographic principle proposed by Fischler $\&$ Susskind \cite{fis1}) represents one of the most studied candidate for DE \cite{fis2,fis3,fis4,fis5}. It was shown by Cohen et al. \cite{15} that, in Quantum Field Theory (QFT), the ultraviolet (UV) cut-off $\Lambda_{UV}$ must be connected to the infrared (IR) one, indicated with $L$, due to the limitations given by the formation of a black hole. If the vacuum energy density produced by UV cut-off is given by $\rho_D = \Lambda_{UV}^4$, then the total energy density of a given size $L$ must be less or at least equal to the mass corresponding to the system-size black hole, i.e.:
\begin{eqnarray}
E_D \leq E_{BH} \rightarrow L^3 \rho_D \leq M_p^2 L, \label{1}
\end{eqnarray}
where $M_p = \left( 8\pi G_N  \right)^{-1/2} \approx 10^{18}$GeV indicates the reduced Planck mass (with $G_N$ representing the Newton's gravitational constant). If the largest possible cut-off $L$ is the one which
saturate this inequality, we can derive the expression for the energy density $\rho_D$ of the HDE model as follow:
\begin{eqnarray}
\rho_D = 3\alpha  M_p^2 L^{-2}, \label{2}
\end{eqnarray}
with $\alpha$ representing a dimensionless constant parameter. The HDE model based on the entropy bound can be also derived with a different approach  \cite{8!!!}. In the black hole thermodynamics \cite{9!!!,9a!!!}, a maximum value of the entropy in a box with a dimension of $L$ (which is also referred as Bekenstein-Hawking entropy bound), exists and it is given by $S_{BH} \approx M_p^2 L^2$, which goes as the area $A$ of the box (given by the approximate relation $A \approx L^2$) rather than its volume $V$ (which goes as $V \approx L^3$). Furthermore, for a macroscopic system with some self-gravitation effects which we can not ignore, the expression of the Bekenstein entropy bound (indicated with $S_B$) is obtained multiplying the energy $E \approx \rho_DL^3$ and the linear size of the system, given by $L$. If we require that the Bekenstein entropy bound is smaller than the Bekenstein-Hawking entropy (i.e. $S_B \leq S_{BH}$, which implies that $E\cdot L \leq M_p^2 L^2$), it is possible to obtain the same result obtained from energy bound argument, i.e. $\rho_D \leq M_p^2L^{-2}$.\\
The HDE model has been accurately investigated in scientific literature in many different ways. Chen et al. \cite{A73} considered the HDE model in order to produce an inflationary epoch in the early evolutionary phases of universe. Jamil et al. \cite{11} studied the equation of state (EoS) parameter $\omega_D$ of the HDE model considering a time-varying Newton's gravitational constant, i.e. $G_N \equiv G_N \left( t \right)$; moreover, the Authors have shown that $\omega_D$ can be significantly modified in the low-redshift limit. \\
The HDE model was recently studied in other recent papers \cite{12a,12b,12d,12g,13a,13b,13c}  with different IR cut-offs, for example the particle horizon, the future event horizon and the Hubble horizon. Moreover, correspondences between the HDE model and some scalar field models have been recently proposed \cite{14,14a,14b}. It was also demonstrated that the HDE model can fit well cosmological data obtained thanks to observations of both CMB radiation anisotropies and SNeIa  \cite{16a,16d,16e}.\\
Recently, the cosmic accelerated expansions has been also accurately studied by imposing the concept of modification of gravity  \cite{odi3,nojo2}. This new model of gravity (predicted by string/M theory) gives a very natural gravitational alternative for exotic matter. The explanation of the phantom, non-phantom and quintom phases of the universe can be well described using modified gravity without introducing a negative kinetic term in DE models.\\
The HDE model has been also tested and constrained by various astronomical observations \cite{A180,A157}  and it has been extended to various physical contexts \cite{26,A81,A92,A208,A166,A213,A142,A215}.\\
The importance of modified gravity models for the late acceleration of the universe has been recently studied by many authors. Some of the most famous and known models of modified gravity are represented by braneworld models, $f\left(T\right)$ gravity (where $T$ indicates the torsion scalar),
$f \left(R\right)$ gravity (where $R$ indicates the Ricci scalar curvature), $f \left(G\right)$ gravity (where $G=R^2-4R_{\mu \nu}R^{\mu \nu} + R_{\mu \nu \lambda \sigma}R^{\mu \nu \lambda \sigma}$ represents the Gauss-Bonnet invariant, with $R$ representing the Ricci scalar curvature, $R_{\mu \nu}$ representing the Ricci curvature tensor and $R_{\mu \nu \lambda \sigma}$ representing the Riemann curvature tensor), $f \left(R,T\right)$ gravity, DGP model, DBI models and Brans-Dicke gravity \cite{odi4,odi5,15gau,15a,15b,15c,15d,15e,15g,15h,15i,15l,mio,miobd1,miobd2,miobd3,dgp1,miodbi,frt2,frt4,frt5,miofg1,miofg2,fr1,fr4,fr5,miofr,fr6,fr7,fr8,fr10,fr12,fr17,
mioft1,mioft2,mioft3,ft2,ft4,ft6,ft7,bra1,bra2,miors}.\\
Ho\v{r}ava \cite{23,A22,A176} recently proposed a new theory of gravity renormalizable with higher spatial derivatives in four dimensions which reduces to Einstein's gravity with non-vanishing value of the cosmological constant in IR but with some improved UV behaviors. Ho\v{r}ava gravity is also similar to a scalar field theory of Lifshitz \cite{lif1} in which the temporal dimension has weight equal to three if the space dimension has weight equal to one. For this reason, the theory introduced by Ho\v{r}ava is also known as Ho\v{r}ava-Lifshitz gravity. Ho\v{r}ava-Lifshitz gravity has been studied and extended in detail \cite{A148,A171,A192,A146,A202,A141,A155,A195,A199} and applied as a cosmological framework of the universe  \cite{46,76,A207,A126,A200,A151,A170,A188,A183,A163,A187,A172,A190,A203,A36,A7}.\\
Ho\v{r}ava-Lifshitz theory is not Lorentz invariant (except in the infrared limit), it is non-relativistic, it leads to the fact that the speed of light $c$ diverges in the ultraviolet (UV) limit and test particles do not follow geodesics. In consequence causal structures are different from those in General Relativity \cite{20} and it is not possible to define the entropy.
There are four different versions of the Ho\v{r}ava-Lifshitz theory:
(i) with projectability condition, (ii) without projectability condition, (iii) with detailed balance and (iv) without detailed balance. At a first look it seems that this non-relativistic model for quantum gravity has a well defined IR limit and it reduces to General Relativity. But as it was first indicated by Mukohyama \cite{57,58}, Ho\v{r}ava-Lifshitz theory mimics General Relativity plus DM. For reviews on the scenario where the cosmological evolution is governed by Ho\v{r}ava-Lifshitz gravity see \cite{59,63,53,74}.\\
Due to these novel features, a huge effort in examining, improving and  extending the properties of the theory itself have been made \cite{A135,A112,A113,73}. Moreover, application of Ho\v{r}ava-Lifshitz gravity as a cosmological framework gives rise to Ho\v{r}ava-Lifshitz cosmology, which leads to interesting behaviors. In particular, it is possible to examine some specific solution subclasses \cite{mina}, the perturbation spectrum \cite{A221}, gravitational wave production \cite{A131}, the matter bounce \cite{A69}, the black hole properties \cite{A78,A110,A65}, the DE phenomenology \cite{app} and the astrophysical phenomenology \cite{kai}.
Ho\v{r}ava-Lifhsitz cosmology has been recently studied using different infrared cut-offs.
For example, Karami et al. \cite{suracon1} studied the logarithmic entropy-corrected New Agegraphic Dark Energy (NADE) model in the framework of Ho\v{r}ava-Lifshitz cosmology. Jamil et al. \cite{suracon3} studied the generalized second law of thermodynamics in Ho\v{r}ava-Lifshitz cosmology using as IR cut-off the dynamical apparent horizon. Karami et al. \cite{suracon2} considered the power-law NADE model in Ho\v{r}ava-Lifshitz cosmology. Jamil et al. \cite{suracon4} worked on the NADE model in Ho\v{r}ava-Lifshitz cosmology. Setare $\&$ Jamil \cite{suracon5} studied the HDE model with varying gravitational constant $G$ in Ho\v{r}ava-Lifshitz cosmology. Pasqua $\&$ Chattopadhyay \cite{suracon6} studied the properties of the logarithmic entropy corrected version of the HDE model in the framework of Ho\v{r}ava-Lifshitz cosmology. Elizalde et al. \cite{odi7} made a work devoted to the unification of inflation with DE in the framework of modified $f\left(R\right)$ Ho\v{r}ava-Lifshitz gravity. Lopez-Revelles et al. \cite{rat1} studied the properties of the ekpyrotic universes in the framework of $f\left(R\right)$ Ho\v{r}ava-Lifshitz gravity. Nugmanova et al. \cite{rat3} studied the cosmological aspects of Ho\v{r}ava-Lifshitz gravity for integrable and nonintegrable models.
However, despite this extended research, there are still many ambiguities if Ho\v{r}ava-Lifshitz gravity is a reliable theory and capable of a successful description of the cosmological behavior of the universe.\\
It must be underlined here that the black hole entropy $S$ assumes an important role in the derivation of HDE energy density. Indeed, we know that the derivation of the HDE model energy density strongly depends on the entropy-area relation given, in Einstein's theory of gravity, by $S \approx A \approx L^2$ (with $A$ indicating the area of the black hole horizon). However, the definition of the entropy-area relation can be modified taking into account some quantum effects, which have their motivation from the Loop Quantum Gravity (LQG). The relation $S\left(A\right)$ has two interesting corrections, i.e. the logarithmic correction \cite{hei17} and power-law correction \cite{das18,das18a,das18b,das18c,das18d,das18e}, which arise in dealing with the entanglement of quantum fields in and out the horizon. \\
The power-law corrected expression of the entropy has the following form:
\begin{eqnarray}
  S\left(A\right)= \frac{A}{4G_N}\left( 1-K_{\alpha}A^{1-\alpha /2}   \right), \label{1}
\end{eqnarray}
where $\alpha$ represents a dimensionless constant whose value is still under debate and not yet clearly determined. Moreover, $K_{\alpha}$ is a constant parameter defined as follow:
\begin{eqnarray}
  K_{\alpha} = \frac{\alpha \left( 4\pi  \right)^{\alpha /2 -1}}{\left( 4-\alpha  \right)r_c^{2-\alpha}}. \label{2}
\end{eqnarray}
In Eq. (\ref{2}), $r_c$ represents the cross-over scale. Furthermore, $A=4\pi R_h^2$ gives the area of the horizon (with $R_h$ being the radius of the horizon). The second term in Eq. (\ref{1}) gives the power-law correction to the entropy-area law. In order to have the entropy as a well-defined quantity, we need that the condition $\alpha>0$ must be satisfied. Motivated by the relation given in Eq. (\ref{1}), a new version of HDE (known as Power Law Entropy Corrected HDE (PLECHDE)) was recently proposed as follow:
\begin{eqnarray}
\rho_D = 3\alpha M_p^2L^{-2} - \varepsilon M_p^2 L^{-\delta}, \label{3}
\end{eqnarray}
where $\delta$ represents a positive power law index and $\alpha$ and $\varepsilon$ are two positive constant parameters.\\
In the limiting case corresponding to $\varepsilon =0$, Eq. (\ref{3}) recovers the well-known HDE energy density. The HDE model can be also recovered, from a mathematical point of view, in the limiting case $\delta \rightarrow \infty$. The correction term in Eq. (\ref{3}) can be comparable to the first one only when $L$ is very small. Hence, at the very early stage (when the universe undergoes an inflationary phase), the correction term in the PLECHDE density becomes important but, when the universe becomes larger, the PLECHDE model reduces to the ordinary HDE model. We need to also note that, after the end of the inflationary phase, the universe subsequently enters in the radiation and then matter dominated eras. In these two epochs, since the universe is much larger, the power-law entropy-corrected term present in the PLECHDE model, namely the second term in Eq. (\ref{3}), can be safely ignored. Therefore, the PLECHDE model can be considered as a model of entropic cosmology which unifies the early-time inflation and late-time cosmic acceleration of the universe.\\
The logarithmic corrected entropy has the following form:
\begin{eqnarray}
S\left(A\right)= \frac{A}{4G_N} + \bar{\alpha} \ln\left(\frac{A}{4G_N}\right) + \bar{\beta}, \label{4}
\end{eqnarray}
where $\bar{\alpha}$ and $\bar{\beta}$ are two dimensionless constant parameters which values are not yet accurately determined.  Motivated by the logarithmic entropy-corrected relation given in Eq. (\ref{4}), the energy density $\rho_D$ of the so-called Logarithmic Entropy Corrected HDE (LECHDE) model can be defined as follow:
\begin{eqnarray}
    \rho_D=3\alpha  M_p^2L^{-2}+ \theta L^{-4} \log \left( M_p^2L^2 \right) + \delta L^{-4},\label{5}
\end{eqnarray}
where $\theta$ and $\delta$ represent two dimensionless constant parameters. In the limiting case corresponding to $\theta = \delta = 0$, Eq. (\ref{5}) recovers the expression of the well-known HDE energy density, as we also obtained in the limiting case for the PLECHDE model. The last two terms in Eq. (\ref{5})  are of the same order of the first one only when $L$ is small, then the corrections given by the extra terms assume a physical meaning only at early evolutionary stages of the universe as we also have for the PLECHDE model. \\
This paper differs from the other ones previously described since we propose to consider as infrared (IR) cut-off the average radius of Ricci scalar curvature, i.e. $L=R^{-\frac{1}{2}}$, for both power law and logarithmic entropy corrected versions of the HDE model. For a non-flat FLRW universe, the Ricci scalar $R$ is given by:
\begin{equation}
 R=6\left(\dot{H}+2H^2+\frac{k}{a\left(t\right)^2}\right),\label{6}
\end{equation}
where $H=\frac{\dot{a}}{a}$ represents the Hubble parameter, $\dot{H}$ is the first derivative of the Hubble parameter with respect to the cosmic time $t$, $a\left(t\right)$ is a dimensionless scale factor (which is function of the cosmic time $t$) and $k$ is the curvature parameter which has dimension of $length^{-2}$ and can assume the values $-1$ $0$ and $+1$ which yield, respectively, a closed, a flat or an open FLRW universe, describing in this way the spatial geometry of space-time.\\
Gao et al. \cite{90} were the first authors to consider the average radius of the Ricci scalar curvature as possible IR cut-off. They derived that, if the future event horizon is considered as infrared cut-off for the HDE model, it leads to the causality problem. For this reason, in order to find a model which can avoid the causality problem, they proposed the average radius of the Ricci scalar curvature of FLRW metric as a new cut-off since this model (known as \textit{Ricci Dark Energy} (\textit{RDE}) model) can resolve not only the causality problem but also the coincidence one. Gao et al. also obtained that, in the limiting case corresponding to $\alpha\simeq0.46$, the RDE model (in the case of absence of entropy corrections) yields the correct DE density and the correct equation of state parameter at present time. \\
We must here underline that the RDE model belongs to class of generalized HDE models which was also introduced in Nojiri $\&$ Odintsov \cite{odi1}. Moreover, thanks to the work proposed by Cai, Hu $\&$ Zhang \cite{91}, which studied the casual entropy bound in the holographic framework, the Ricci model had an appropriate motivation for which it could be studied. Furthermore, the RDE model is compatible with observational data of Supernovae, CMB radiation anisotropies, Baryon Acoustic Oscillations (BAO), gas mass fraction in galaxy clusters, the history of the Hubble function and the growth function \cite{duranduran}. RDE was studied in different ways: in fact, we can find in literature studies of RDE model in Brans-Dicke cosmology \cite{mioBD}, correspondence of the RDE model with scalar fields \cite{mio1,mio2,mio3}, the statefinder diagnostic of RDE \cite{92}, reconstruction of $f\left(R \right)$ \cite{93}, quintom \cite{94}, contributions of viscosity to RDE \cite{95} and related observational constraints \cite{96}. However, there are also some criticisms regarding the RDE model:  Kim et al. \cite{kimkim} pointed out that an accelerating phase of the RDE is the same of a constant DE model, which implies that the RDE model may not be a new model able to explain the present accelerated expansion of the universe.\\
Replacing $L$ with $R^{-1/2}$ in Eqs. (\ref{3}) and (\ref{5}), we get the energy density of the R-PLECHDE and of the R-LECHDE models, respectively, as follow:
\begin{eqnarray}
\rho_D &=& 3\alpha M^2_p R-\varepsilon M^2_p R^{\frac{\alpha}{2}},\label{7} \\
\rho_D &=& 3\alpha M_p^2R+\theta R^2 \log\left(\frac{M_p^2}{R}\right)+\delta R^2. \label{8}
\end{eqnarray}
The main aim of this paper is to study the models given in Eqs. (\ref{7}) and (\ref{8}) in the framework of Ho\v{r}ava-Lifshitz cosmology.\\
This paper is organized as follow. In Section 2, we describe the most important features of Ho\v{r}ava-Lifshitz cosmology. In Section 3, we study the RDE model considered in the context of Ho\v{r}ava-Lifshitz cosmology. Finally, in Section 4 we write the Conclusions of this work.

\section{Ho\v{r}ava-Lifshitz Gravity}
In this Section, we want to describe the main cosmological features of Ho\v{r}ava-Lifshitz cosmology. \\
Under the projectability condition, the full metric in the (3+1)-dimensional Arnowitt-Deser-Misner formalism is given by \cite{arno}:
\begin{eqnarray}
ds^2 = -N^2dt^2+g_{ij}\left(dx^i+N^i dt \right)\left(dx^j+N^jdt \right), \label{9}
\end{eqnarray}
where $t$ refers to the cosmic time and the dynamical variables $N$, $N^i$ and $g_{ij}$ are the lapse function, the shift vector and the 3-dimensional metric tensor, respectively. The projectability condition implies that the lapse function  must be space-independent, while the shift vector $N^i$ and the 3-dimensional metric $g_{ij}$ are still dependent on both space and time and indices are raised and lowered using the metric tensor $g_{ij}$. The scaling transformations of the coordinates are given by:
\begin{eqnarray}
x^i &&\rightarrow l x^i, \\
t&&\rightarrow l^z t, \label{tra}
\end{eqnarray}
where $\l$, $z$, $x$ and $t$  represent, respectively, the scaling factor, the dynamical critical exponent, the spatial coordinates and  the temporal coordinate.\\
In case considered in this paper, we have that $z=3$, then Eq. (\ref{tra}) leads to:
\begin{eqnarray}
t\rightarrow l^3 t.
\end{eqnarray}
We must also underline that $i$ is referred to the three spatial coordinates.\\
The gravitational action $S_g$ of Ho\v{r}ava-Lifshitz cosmology can be decomposed into a kinetic part $L_K$ and a potential part $L_V$ as follow:
\begin{eqnarray}
S_g = \int dt d^3x \sqrt{g} N \left( L_K + L_V   \right),
\end{eqnarray}
where $g$ represents the determinant of the metric tensor $g^{\mu \nu}$.\\
The assumption of detailed balance \cite{A177} reduces the number of possible terms in
the Lagrangian, and it allows for a quantum inheritance principle, since the $(D + 1)$-dimensional
theory acquires the renormalization properties of the $D$-dimensional one. Under the detailed balance condition, the full action $S_g$ of Ho\v{r}ava-Lifshitz gravity can be written as follow \cite{A177}:
\begin{eqnarray}
S_g &=& \displaystyle \int{dtd^3x \sqrt{g}N} \left[\frac{2}{\kappa^2} \left(K_{ij}K^{ij}-\lambda K^2\right) \right.\nonumber\\
&& + \frac{\kappa^2}{2\omega^4}C_{ij}C^{ij} -\frac{\kappa^2\mu}{2\omega^2} \frac{\eta^{ijk}}{\sqrt{g}}R_{il} \bigtriangledown_j R^l_k \nonumber\\
&& +\frac{\kappa^2\mu^2}{8}R_{ij}R^{ij}  \nonumber\\
&& + \left. \frac{\kappa^2\mu^2}{8\left(3\lambda - 1\right)}\left(\frac{1-4\lambda}{4}R^2+\Lambda R - 3 \Lambda^2\right)\right], \label{10}
\end{eqnarray}
where $K_{ij}$ and $C_{ij}$ are the extrinsic curvature and the Cotton tensor which are defined, respectively, as:
\begin{eqnarray}
K_{ij} &=& \frac{1}{2N}\left(\dot{g}_{ij}- \bigtriangledown_iN_j - \bigtriangledown_jN_i\right), \label{11} \\
C_{ij} &=& \frac{e^{ijk}}{\sqrt{g}} \bigtriangledown_k \left(R^j_i - \frac{1}{4}R \delta^j_i\right). \label{12}
\end{eqnarray}
Moreover, $\eta^{ijk}$ represents the totally antisymmetric unit tensor, $\lambda$ is a dimensionless constant parameter and $\Lambda$ represents a positive dimensionless constant which is related to the cosmological constant in the IR limit.\\
The three variables $\kappa$, $\omega$ and $\mu$ are constants which have mass dimensions of -1, 0 and 1, respectively. \\
In order to include the matter component in a universe governed by Ho\v{r}ava gravity, two option can be considered. In the first one, we introduce a scalar field $\phi$ which action $S_{\phi}$ is given by \cite{12calcagni}:
\begin{eqnarray}
S_m \equiv S_{\phi} &=& \int dtd^3x \sqrt{g}N \left[ \frac{3\lambda -1}{4}\frac{\dot{\phi}^2}{N^2} +m_1m_2 \phi \nabla ^2 \phi \right. \nonumber \\
&&\left. - \frac{1}{2}m_2^2\phi \nabla ^4 \phi + \frac{1}{2}m_3^2\phi \nabla ^6 \phi - V\left( \phi  \right)  \right],
\end{eqnarray}
where $m_i$ are constants and $V\left( \phi  \right) $ is the potential term. Moreover, the equation of motion for the field $\phi$ is given by:
\begin{eqnarray}
\ddot{\phi} + 3H\dot{\phi}+ \frac{2}{3\lambda -1} \frac{dV\left( \phi  \right)}{d \phi} = 0, \label{medioman}
\end{eqnarray}
where $\ddot{}$ indicates a double time derivative while $\dot{}$ a single time derivative. Obviously, in Eq. (\ref{medioman}) we must have that $3\lambda -1 \neq 0$.\\
The second way to insert the matter component is made considering a hydrodynamical approximation and add a cosmological stress-energy tensor to the gravitational field equations, with the condition that the general relativistic formalism must be recovered in the low-energy limit \cite{14cha}. In this case, the pressure $p_m$ and the energy density $\rho_m$ of DM must obey the following continuity equation:
\begin{eqnarray}
\dot{\rho}_m +3H\left(\rho_m+p_m\right) = 0. \label{13}
\end{eqnarray}
In this work, we consider the hydrodynamical approximation.\\
It is well-known in scientific community that Eq. (\ref{10}) has several problems, for example instability, inconsistency and strong coupling problems \cite{57}. In order to overcome them, we invoke the Vainshtein mechanism, as it was already made by Mukohyama for spherical space-times \cite{57} and by Wang $\&$ Wu in the cosmological setting \cite{75}. These considerations were further carried out by using the so-called gradient expansion method \cite{29}. Another approach which can be considered is to introduce an extra $U\left(1\right)$ symmetry, as it was done for the first time by Ho\v{r}ava $\&$ Melby-Thompson \cite{24} with $\lambda =1$, and later on generalized to the case with any $\lambda$ by da Silva \cite{17} (another important paper which needs to be cited in this environment is \cite{odi6)}. These works were further generalized to the case corresponding to absence of the projectability condition \cite{89}. In both cases (with and without the projectability condition), the spin-0 gravitons are eliminated (due to the $U\left(1\right)$ symmetry) and all the problems related to them  are then resolved.\\
In the cosmological context, we use a FLRW metric which is obtained when:
\begin{eqnarray}
N &=& 1, \label{14} \\
 g_{ij}&=&a^2\left(t\right)\gamma_{ij},\label{15} \\
N^i&=&0,  \label{16}
\end{eqnarray}
where $\gamma_{ij}$ is defined as follow:
\begin{eqnarray}
\gamma_{ij}dx^idx^j=\frac{dr^2}{1-kr^2}+r^2d\Omega^2_2. \label{17}
\end{eqnarray}
The differential term $d\Omega^2_2$ indicates the angular part of the metric.\\
Varying the action $S_g$ given in Eq. (\ref{10}), respectively, with respect to the metric components $N$ and $g_{ij}$, we derive the modified Friedmann equations in the context of Ho\v{r}ava-Lifshitz cosmology as follow:
\begin{eqnarray}
H^2 &=& \frac{\kappa^2}{6\left(3\lambda-1\right)}\rho_m + \frac{\kappa^2}{6\left(3\lambda-1\right)}\left[\frac{3\kappa^2\mu^2k^2}{8\left(3\lambda-1\right)a^4} \right. \nonumber\\
&& + \left. \frac{3\kappa^2\mu^2\Lambda^2}{8\left(3\lambda-1\right)}\right] - \frac{\kappa^4\mu^2\Lambda k}{8\left(3\lambda - 1\right)^2a^2}\label{18} , \\
\dot{H}+\frac{3}{2}H^2 &=& - \frac{\kappa^2}{4\left(3\lambda - 1\right)}p_m \nonumber\\
&& -  \frac{\kappa^2}{4\left(3\lambda - 1\right)}\left[\frac{\kappa^2\mu^2k^2}{8\left(3\lambda-1\right)a^4} - \frac{3\kappa^2\mu^2\Lambda^2}{8\left(3\lambda-1\right)} \right] \nonumber\\
&& - \frac{\kappa^4\mu^2\Lambda k}{16\left(3\lambda - 1\right)^2a^2}. \label{19}
\end{eqnarray}
In the limiting case corresponding to the curvature parameter $k$ equal to zero, i.e. $k = 0$, the higher order derivative terms give no contribution to the action. Instead, when $k \neq 0$, the higher derivative terms become relevant for small volumes, i.e. for small values of the scale factor $a$, and become insignificant for large values of the scale factor $a$, where it agrees with General Relativity. \\
Using the Friedmann equations obtained in Eqs. (\ref{18}) and (\ref{19}), we can define the energy density $\rho_D$ and the pressure $p_D$ of DE respectively as follow:
\begin{eqnarray}
\rho_D &\equiv& \frac{3\kappa^2\mu^2k^2}{8\left(3\lambda-1\right)a^4} +  \frac{3\kappa^2\mu^2\Lambda^2}{8\left(3\lambda-1\right)}, \label{20} \\
p_D &\equiv& \frac{\kappa^2\mu^2k^2}{8\left(3\lambda-1\right)a^4} -  \frac{3\kappa^2\mu^2\Lambda^2}{8\left(3\lambda-1\right)}. \label{21}
\end{eqnarray}
The first term on the right hand side of Eqs. (\ref{20}) and (\ref{21}) (which is proportional to $a^{-4}$) represents effectively the dark radiation term present in Ho\v{r}ava-Lifshitz cosmology, while the second term (which is a constant term) present on both equations behaves like a cosmological constant term.
Moreover, Eqs. (\ref{20}) and (\ref{21}) satisfy the following continuity equation:
\begin{eqnarray}
\dot{\rho}_D +3H\left(\rho_D+p_D\right) = 0. \label{22}
\end{eqnarray}
Furthermore, Eqs. (\ref{18}) and (\ref{19}) reduce to the standard Friedmann equations $\left(c = 1\right)$ if we consider:
\begin{eqnarray}
G_{cosmo} = \frac{\kappa^2}{16\pi\left(3\lambda-1\right)}, \label{23} \\
\frac{\kappa^4\mu^2\Lambda}{8\left(3\lambda-1\right)^2} = 1, \label{24}
\end{eqnarray}
where $G_{cosmo}$ represents the Newton's cosmological constant. We must here emphasize that, in gravitational theories with the violation of Lorentz invariance (like Ho\v{r}ava-Lifshitz cosmology), the Newton's gravitational constant $G_{grav}$ (which is present in the gravitational action $S_g$) differs with the Newton's cosmological constant $G_{cosmo}$ (which
is present in Friedmann equations) unless Lorentz invariance is restored.\\
For completeness, we now define $G_{grav}$ as follow:
\begin{eqnarray}
G_{grav} = \frac{\kappa^2}{32\pi}, \label{25}
\end{eqnarray}
as it can be easily derived from Eq. (\ref{10}). We can easily see that, in the infrared (IR) limit (corresponding to $\lambda = 1$), where Lorentz invariance is restored, $G_{cosmo}$ and $G_{grav}$ are equivalent.\\
Furthermore, using Eqs. (\ref{20}), (\ref{21}), (\ref{23})  and (\ref{24}), we can rewrite the modified Friedmann equations given in Eqs. (\ref{18}) and (\ref{19}) in their usual forms:
\begin{eqnarray}
H^2+\frac{k}{a^2} &=&  \frac{8\pi G_{cosmo}}{3}\left(\rho_m+\rho_D \right), \label{26}\\
\dot{H}+ \frac{3}{2}H^2+\frac{k}{2a^2} &=& -4\pi G_{cosmo}  \left(p_m+p_D \right).  \label{27}
\end{eqnarray}

\section{Ricci Dark Energy Models in Ho\v{r}ava-Lifshitz Cosmology}
We now want to investigate the properties of the Ricci Dark Energy (RDE) models in the framework of Ho\v{r}ava-Lifhsitz cosmology. We consider a spatially non-flat FLRW universe containing DE and DM.\\
The dimensionless fractional energy densities for DM, DE and curvature parameter $k$ are defined, respectively, as follow:
\begin{eqnarray}
\Omega_m &=& \frac{\rho_m}{\rho_{cr}}=\frac{8\pi G_{cosmo}}{3H^2}\rho_m, \label{28} \\
\Omega_D &=& \frac{\rho_D}{\rho_{cr}}=\frac{8\pi G_{cosmo}}{3H^2}\rho_D, \label{29}  \\
\Omega_k &=& -\frac{k}{a^2H^2}, \label{30}
\end{eqnarray}
where $\rho_{cr}$ represents the critical energy density defined as:
\begin{eqnarray}
\rho_{cr} = \frac{3H^2}{8\pi G_{cosmo}}. \label{31}
\end{eqnarray}
Using Eqs. (\ref{28}), (\ref{29}) and (\ref{30}), the first Friedmann equation given in Eq. (\ref{26}) can be rewritten as:
\begin{eqnarray}
1 - \Omega_k = \Omega_D + \Omega_m . \label{32}
\end{eqnarray}
We now want to derive the expression of the EoS parameter $\omega_D$ for the models considered in this paper.
Using the Friedmann equation given in Eq. (\ref{26}), the Ricci scalar $R$ can be  also rewritten as:
\begin{eqnarray}
    R=6\left[  \dot{H} + H^2 + \frac{8\pi G_{cosmo}}{3}\left(\rho_m+\rho_D\right) \right].\label{33}
\end{eqnarray}
Our goal is now to derive an expression for the quantity $ \dot{H} + H^2$.\\
Differentiating the Friedmann equation given in Eq. (\ref{26}) with respect to the cosmic time $t$, we derive the following expression for $\dot{H}$:
\begin{eqnarray}
    \dot{H}=\frac{k}{a^2}-4\pi G_{cosmo}\left[  \rho_m+\rho_D\left( 1 + \omega_D \right) \right],\label{34}
\end{eqnarray}
where $\omega_D = p_D / \rho_D$ represents the EoS parameter of DE.\\
Adding Eqs. (\ref{26}) and (\ref{34}), we obtain that the term $\dot{H} + H^2$ can be written as:
\begin{eqnarray}
    \dot{H}+H^2=-\frac{4\pi G_{cosmo}}{3}\left[  \rho_m+\rho_D\left( 1 + 3\omega_D \right) \right].\label{35}
\end{eqnarray}
Therefore, using Eq. (\ref{35}), the Ricci scalar $R$ given in Eq. (\ref{33}) can be rewritten as follow:
\begin{eqnarray}
 R=8\pi G_{cosmo}\left(\rho_m+\rho_D\right)- 24\pi G_{cosmo}\rho_D\omega_D.\label{36}
\end{eqnarray}
The EoS parameter $\omega_D$ can be now easily obtained from Eq. (\ref{36}) as follow:
\begin{eqnarray}
 \omega_D &=& -\frac{R}{24\pi G_{cosmo} \rho_D}+ \frac{\rho_D+\rho_{m}}{3\rho_D} = \nonumber \\
 && -\frac{R}{24\pi G_{cosmo} \rho_D} + \frac{\Omega_D+\Omega_{m}}{3\Omega_D} = \nonumber \\
 && -\frac{R}{24\pi G_{cosmo} \rho_D} + \frac{1- \Omega_{k}}{3\Omega_D} , \label{37}
\end{eqnarray}
where we used the relation $\frac{\rho_D+\rho_m}{3\rho_D} =\frac{\Omega_D+\Omega_m}{3\Omega_D}= \frac{1-\Omega_k}{3\Omega_D}$ along with Eq. (\ref{32}).\\
The final expressions of the equation of state parameter $\omega_D$ for the model considered in this work can be obtained substituting in Eq. (\ref{37}) the energy densities of DE given in Eqs. (\ref{7}) and (\ref{8}).

\subsection{Non-interacting Case}
We start considering the case corresponding to absence of interaction between DE and DM.\\
We consider a FLRW universe filled with DE and DM, which is considered pressureless (i.e., $p_m =0$), evolving according
to conservation laws which can be expressed by the following continuity equations:
\begin{eqnarray}
\dot{\rho}_D +3H\left(1+\omega_D\right)\rho_D = 0, \label{38} \\
\dot{\rho}_m +3H\rho_m = 0. \label{39}
\end{eqnarray}
From Eq. (\ref{38}), we can easily obtain the following expression for the time derivative of the DE energy density $\dot{\rho}_D$:
\begin{eqnarray}
    \dot{\rho}_D = -3H \rho_D\left( 1+\omega_D  \right). \label{40}
\end{eqnarray}
Using the expression of $\omega_D$ given in Eq. (\ref{37}), we can write Eq. (\ref{40}) as:
\begin{eqnarray}
    \dot{\rho}_D=3H\left[ -\rho_D-\frac{\rho_m + \rho_D}{3}  + \frac{R}{24\pi G_{cosmo} } \right].\label{41}
\end{eqnarray}
Dividing Eq. (\ref{41}) by the critical energy density $\rho_{cr}$ defined in Eq. (\ref{31}), we obtain:
\begin{eqnarray}
    \frac{\dot{\rho}_D}{\rho_{cr}}=3H\left[ -\Omega_D-\frac{1 - \Omega_k}{3}  + \frac{R}{9H^2} \right]=\dot{\Omega}_D+2\Omega_D\frac{\dot{H}}{H},  \label{42}
\end{eqnarray}
where we used Eq. (\ref{32}).\\
 Using the definition of the Ricci scalar curvature $R$ given in Eq. (\ref{6}), we obtain that the term $\frac{R}{9H^2}$ is equivalent to:
\begin{eqnarray}
    \frac{R}{9H^2}=\frac{2}{3}\left( \frac{\dot{H}}{H^2} +2 - \Omega_k \right).\label{43}
\end{eqnarray}
Substituting Eq. (\ref{43}) in Eq. (\ref{42}), we obtain:
\begin{eqnarray}
\dot{\Omega}_D=2\frac{\dot{H}}{H}\left(1-\Omega_D  \right)+3H\left[-\Omega_D + \frac{1-\Omega_k}{3} + \frac{2}{3} \right].\label{44}
\end{eqnarray}
Since  $\Omega_D'=\frac{d\Omega_D}{dx}= \frac{1}{H}\dot{\Omega}_D$ (where $x=\ln a$, a prime indicates a derivative with respect to $x$ while a dot indicates a derivative with respect the cosmic time $t$), we can write:
\begin{eqnarray}
H   \Omega_D'=2H'\left(1-\Omega_D  \right)+3H\left[-\Omega_D +\frac{1 - \Omega_k}{3} + \frac{2}{3} \right],\label{45}
\end{eqnarray}
which yields to:
\begin{eqnarray}
    \Omega_D'&=&\frac{2}{H}\left(1-\Omega_D  \right)+3\left[-\Omega_D +\frac{1 - \Omega_k}{3} + \frac{2}{3} \right] = \nonumber \\
      &&\left( 1-\Omega_D - \Omega_k  \right) +2\left( 1-\Omega_D   \right)\left( 1 + \frac{1}{H}  \right).   \label{46}
\end{eqnarray}
In Eq. (\ref{46}), we used the relation:
\begin{eqnarray}
 \frac{\dot{H}}{H}=H'=\frac{a'}{a}=1. \label{47}
\end{eqnarray}
Differentiating the expression of $\Omega_k$ given in Eq. (\ref{30}) with respect to $x= \ln a$, we obtain:
\begin{eqnarray}
\Omega'_k = -2 \Omega_k \left(  1+ \frac{1}{H}  \right), \label{48}
\end{eqnarray}
where we used the relation given in Eq. (\ref{47}). \\
The EoS parameter $\omega_D$ of the DE can be also parameterized as function of the redshift $z$ as follow \cite{28}:
\begin{eqnarray}
\omega_D\left(z\right) = \omega_0 + \omega_1 z. \label{49}
\end{eqnarray}
We must remember that the relation between redshift $z$ and scale factor $a$ is given by:
\begin{eqnarray}
a = \frac{1}{1+z} = \left( 1+z  \right)^{-1}  \rightarrow z= a^{-1}-1.     \label{50}
\end{eqnarray}
Using Eqs. (\ref{38}) and (\ref{49}), we obtain that the DE energy density evolves as follow \cite{84}:
\begin{eqnarray}
\frac{\rho_D}{\rho_{D_0}} = a^{-3 \left( 1+ \omega_0 - \omega_1 \right)} e^{3\omega_1z}. \label{51}
\end{eqnarray}
The Taylor expansion of the DE energy density $\rho_D$ around $a_0 = 1$ gives:
\begin{eqnarray}
\ln{\rho_D} = \ln{\rho_{D_0}} + \left. \frac{d\ln{ \rho_D}}{d\ln{ a}}\right|_0\ln{a} + \left. \frac{1}{2} \frac{d^2\ln{ \rho_D}}{d\left(\ln{ a}\right)^2}\right|_0\left(\ln{a}\right)^2 + ... , \label{52}
\end{eqnarray}
where the index 0 denotes the value of a quantity at present time (with this notation, $a_0$ indicates the present day value of the scale factor $a\left(t \right)$ while $\rho_{D_0}$ indicates the present day value of the energy density of DE $\rho_D$). Using Eq. (\ref{50}), we can write, for small redshifts, the following expansion of the scale factor:
\begin{eqnarray}
\ln a = -\ln \left( 1+z  \right) \simeq -z + \frac{z^2}{2}. \label{53}
\end{eqnarray}
Then, Eqs. (\ref{51}) and (\ref{52}) reduce, respectively, to:
\begin{eqnarray}
\frac{\ln{\left(\rho_D/\rho_{D_0}\right)}}{\ln{a}} &=& -3\left(1+\omega_0\right) - \frac{3}{2}\omega_1z, \label{54} \\
\frac{\ln{\left(\rho_D/\rho_{D_0}\right)}}{\ln{a}} &=& \left. \frac{d\ln{ \rho_D}}{d\ln{ a}}\right|_0 - \left.  \frac{1}{2} \frac{d^2\ln{ \rho_D}}{d\left(\ln{ a}\right)^2}\right|_0 z. \label{55}
\end{eqnarray}
Comparing Eqs. (\ref{54}) and (\ref{55}), we obtain the following expressions for the two parameters $\omega_0$ and $\omega_1$:
\begin{eqnarray}
\omega_0  &=& \left. -\frac{1}{3}\frac{d\ln{ \rho_D}}{d\ln{ a}}\right|_0 - 1, \label{56} \\
\omega_1  &=& \left. \frac{1}{3}\frac{d^2\ln{ \rho_D}}{d\left(\ln{ a}\right)^2}\right|_0. \label{57}
\end{eqnarray}
From Eq. (\ref{39}), we obtain that the energy density of DM $\rho_m$ evolves as $\rho_m = \rho_{m_0}a^{-3}$, where $\rho_{m_0}$ represents the present day value of $\rho_m$. Using Eq. (\ref{32}), we get:
\begin{eqnarray}
\rho_D = \left(\frac{\rho_m}{\Omega_m}\right)\Omega_D = \frac{\rho_{m_0}a^{-3}}{\left(1 - \Omega_k - \Omega_D\right)}\Omega_D, \label{58}
\end{eqnarray}
which, substituted  into Eq. (\ref{56}), yields:
\begin{eqnarray}
\omega_0 = -\frac{1}{3} \left[\frac{\Omega'_D}{\Omega_D} + \frac{\Omega'_D + \Omega'_k}{\left(1 - \Omega_k - \Omega_D\right)} \right]_0. \label{59}
\end{eqnarray}
Instead, inserting Eq. (\ref{58}) into Eq. (\ref{57}), we obtain the following relation for $\omega_1$:
\begin{eqnarray}
\omega_1= \frac{1}{3} \left[\frac{\Omega''_D}{\Omega_D} - \frac{\Omega'^2_D}{\Omega^2_D} + \frac{\Omega''_D + \Omega''_k}{\left(1 - \Omega_k - \Omega_D\right)} + \frac{ \left( \Omega'_D + \Omega'_k \right)^2}{\left(1 - \Omega_k - \Omega_D\right)^2} \right]_0.
 \label{60}
\end{eqnarray}
We now calculate some useful quantities in order to obtain final expressions of $\omega_0$ and $\omega_1$.\\
Adding Eqs. (\ref{46}) and (\ref{48}), we obtain, after some algebraic calculations, that:
\begin{eqnarray}
\Omega_D' + \Omega_k' =  \left( 1-\Omega_D - \Omega_k  \right)\left(  3+ \frac{2}{H}  \right). \label{61}
\end{eqnarray}
We also derive that the quantities $\frac{\Omega_D'}{\Omega_D}$ and $\frac{\Omega_k'}{\Omega_D}$ are given, respectively, by:
\begin{eqnarray}
 \frac{\Omega_D'}{\Omega_D} &=& \frac{\left( 1-\Omega_D - \Omega_k  \right)}{\Omega_D} +2\left( \frac{1-\Omega_D}{\Omega_D}   \right)\left( 1 + \frac{1}{H}  \right),   \label{62} \\
\frac{\Omega_k'}{\Omega_D} &=& -\frac{2\Omega_k}{\Omega_D}\left( 1+\frac{1}{H}  \right). \label{63}
\end{eqnarray}
Taking the derivative  of Eq. (\ref{48}) with respect to $x= \ln a$, we obtain:
\begin{eqnarray}
\Omega''_k = -2 \Omega'_k \left(  1+ \frac{1}{H}  \right) + \frac{2\Omega_k}{H^2}. \label{66}
\end{eqnarray}
Moveover, the derivative with respect to  $x= \ln a$ of Eq. (\ref{46}) leads to:
\begin{eqnarray}
\Omega_D''=  -\Omega'_D\left( 3+\frac{2}{H}   \right) - \frac{2}{H^2}\left(  1-\Omega_D \right)-\Omega'_k. \label{67}
\end{eqnarray}
Adding Eqs. (\ref{66}) and (\ref{67}), we obtain:
\begin{eqnarray}
\Omega_D'' + \Omega_k'' &=&  -\frac{2}{H^2} \left( 1-\Omega_D - \Omega_k  \right) - \left(\Omega_D' + \Omega_k'\right) \left( 3 + \frac{2}{H}  \right) = \nonumber \\
 &&-\frac{2}{H^2} \left( 1-\Omega_D - \Omega_k  \right) - \left( 1-\Omega_D - \Omega_k  \right)\left( 3 + \frac{2}{H}  \right)^2 = \nonumber \\
 &&-\left( 1-\Omega_D - \Omega_k  \right) \left[\frac{2}{H^2} +  \left( 3 + \frac{2}{H}  \right)^2    \right]. \label{68}
\end{eqnarray}
Dividing Eq. (\ref{67}) by $\Omega_D$ yields:
\begin{eqnarray}
 \frac{\Omega_D''}{\Omega_D}= -\frac{\Omega_D'}{\Omega_D}\left( 3+\frac{2}{H}   \right) - \frac{2}{H^2}\left(\frac{  1-\Omega_D}{\Omega_D} \right)-\frac{\Omega_k'}{\Omega_D}. \label{69}
\end{eqnarray}
Using the expressions of $\frac{\Omega_D'}{\Omega_D} $ and $ \frac{\Omega_k'}{\Omega_k} $ given, respectively, in Eqs. (\ref{62}) and (\ref{63}), we can write Eq. (\ref{69}) as:
\begin{eqnarray}
 \frac{\Omega_D''}{\Omega_D}&=&- \frac{2}{H^2}\left(\frac{  1-\Omega_D}{\Omega_D} \right) + \frac{2\Omega_k}{\Omega_D}\left( 1+\frac{1}{H}  \right) - \left( 3+\frac{2}{H}   \right) \frac{\left( 1-\Omega_D - \Omega_k  \right)}{\Omega_D} - \nonumber \\
  &&2\left( 3+\frac{2}{H}   \right) \left( \frac{1-\Omega_D}{\Omega_D}   \right)\left( 1 + \frac{1}{H}  \right). \label{70}
\end{eqnarray}
Inserting Eqs. (\ref{61}) and (\ref{62}) in the expression of $\omega_0$ given in Eq. (\ref{59}), we derive the following expression for $\omega_0$:
\begin{eqnarray}
\omega_0 = -\frac{1}{3}\left[ 1+ \frac{1}{\Omega_{D_0}}\left( 3+ \frac{2}{H_0} - \Omega_{D_0} - \Omega_{k_0}  \right)   \right].\label{64}
\end{eqnarray}
Instead, using Eqs. (\ref{61}), (\ref{62}) and (\ref{70}), the final expression of $\omega_1$ can be written as:
\begin{eqnarray}
\omega_1 &=& \frac{1}{3}\left\{ -\frac{2}{H_0^2 \Omega_{D_0}} + \frac{2\Omega_{k_0}}{\Omega_{D_0}}\left( 1+\frac{1}{H_0}\right)  -  \right. \nonumber \\
&& \left. \left( 3+\frac{2}{H_0}  \right)\left[ \frac{\left( 1-\Omega_{D_0} - \Omega_{k_0}  \right)}{\Omega_{D_0}} + 2 \left( 1+\frac{1}{H_0}\right) \left( \frac{1-\Omega_{D_0}}{\Omega_{D_0}}  \right)  \right] - \right. \nonumber \\
&& \left. - \left[ \frac{\left( 1-\Omega_{D_0} - \Omega_{k_0}  \right)}{\Omega_{D_0}} +  2\left( 1+\frac{1}{H_0}\right) \left( \frac{1-\Omega_{D_0}}{\Omega_{D_0}}  \right)  \right] ^2 \right\}. \label{71}
\end{eqnarray}
%Inserting the values of the relevant cosmological parameters in Eqs. (\ref{64}) and (\ref{71}), we obtain that:
%\begin{eqnarray}
%\omega_0 &=& -1.45,\\
%\omega_1 &=& 0.72.
%\end{eqnarray}

\subsection{Interacting case}
We now extend our work to the case of presence of interaction between the Dark sectors. The presence of interaction causes the energy conservation law for each dark component not to be held separately, i.e.:
\begin{eqnarray}
\dot{\rho}_D +3H\left(1+\Omega_D\right)\rho_D &=& -Q, \label{72} \\
\dot{\rho}_m +3H\rho_m &=& Q. \label{73}
\end{eqnarray}
$Q$ is an interaction term which is function of cosmological parameters, like the Hubble parameter $H$ and energy densities  of DM and DE $\rho_m$ and $\rho_D$. Among the many different candidates recently proposed in order to describe $Q$, we have chosen here to consider the following one \cite{A209}:
\begin{eqnarray}
    Q = 3b^2H \rho_D,\label{74}
\end{eqnarray}
with $b^2$ representing a coupling parameter (also known as transfer strength) between DM and DE \cite{q1,q1-1,q1-3,q1-4,q1-5,q1-7,q1-8,q1-9,q1-10}. The limiting case corresponding to $b^2 = 0$ represents the non-interacting FLRW model.  \\
Thanks to observational data of Gold SNeIa samples, CMB data from
WMAP satellite and the Baryonic Acoustic Oscillations (BAO) from the Sloan Digital Sky Survey (SDSS),
it was possible to estimate that the coupling parameter between DM and DE must assume a small positive value
of the order of unity, which satisfies the requirement for solving the cosmic coincidence problem
and also constraints given by the second law of thermodynamics \cite{feng08,A173,22}.\\
Some constraints on the value of the coupling parameter $b^2$ have been recently obtained \cite{b2}.  A negative value of the coupling parameter is avoided since it leads to violations of thermodynamical laws.   We need also to emphasize that other more general interaction terms can be used \cite{jamil-08-2008}.\\
The interaction between DE and DM can be detected during the formation of the large scale structures (LSS). It was suggested that the dynamical equilibrium of collapsed structures like galaxy clusters (for example Abell A586) would be modified due to the coupling between DE and DM \cite{1,10,9}. The main idea is that the
virial theorem results to be modified by the energy exchange between DE and DM leading to a bias in
the estimation of the virial masses of clusters when the usual virial conditions are employed. This
gives a probe in the near universe of the dark coupling. \\
Following the same procedure of previous Subsection, we find the following expression for $\Omega'_D$:
\begin{eqnarray}
    \Omega_D'=  \left( 1-\Omega_D - \Omega_k  \right) +2\left( 1-\Omega_D   \right)\left( 1 + \frac{1}{H}  \right) -3b^2 \Omega_D. \label{75}
\end{eqnarray}
The expression of $\Omega_k'$ is the same as in Eq. (\ref{48}).\\
From the expression:
\begin{eqnarray}
\rho_D = \left( \frac{\rho_m}{\Omega_m}\right)\Omega_D =  \frac{\rho_m}{\left(1 - \Omega_k - \Omega_D\right)}\Omega_D, \label{76}
\end{eqnarray}
we obtain the following relation:
\begin{eqnarray}
\frac{d\ln{\rho_D}}{d\ln{a}} = \frac{\rho'_m}{\rho_m} - \frac{\Omega'_m}{\Omega_m} + \frac{\Omega'_D}{\Omega_D}. \label{77}
\end{eqnarray}
We also derive, using Eqs. (\ref{49}) and (\ref{72}), that the interacting DE energy density $\rho_D$ evolves as follow:
\begin{eqnarray}
\frac{\rho_D}{\rho_{D_0}} = a^{-3\left(1+\omega_0 - \omega_1+b^2\right)}e^{3\omega_1 z}.
\label{78}
\end{eqnarray}
Using Eq. (\ref{52}) for small redshifts, Eq. (\ref{78}) reduces to:
\begin{eqnarray}
\frac{\ln{\left(\rho_D / \rho_{D_0}\right)}}{\ln{a}} = -3\left(1+\omega_0+b^2\right) - \frac{3}{2} \omega_1 z. \label{79}
\end{eqnarray}
Comparing Eqs. (\ref{78}) and (\ref{79}), we derive the following expressions for the parameters $\omega_0$ and $\omega_1$  for the interacting DE and DM:
\begin{eqnarray}
\omega_0 &=& \left. -\frac{1}{3}\frac{d\ln{\rho_D}}{d\ln{a}}\right|_0 - 1 - b^2, \label{80} \\
\omega_1 &=& \left. \frac{1}{3}\frac{d^2\ln{\rho_D}}{d \left(\ln{a}\right)^2}\right|_0. \label{81}
\end{eqnarray}
Substituting Eq. (\ref{77}) into Eq. (\ref{80}) and using Eq. (\ref{73}), we can write the parameter $\omega_0$ as:
\begin{eqnarray}
\omega_0 = -\frac{1}{3}\left[\frac{\Omega'_D}{\Omega_D}+\frac{\Omega'_D + \Omega'_k}{\left(1 - \Omega_k - \Omega_D\right)} \right]_0-b^2\left(\frac{1-\Omega_k}{1 - \Omega_k - \Omega_D}\right)_0. \label{82}
\end{eqnarray}
Furthermore, using Eqs. (\ref{72}) and (\ref{81}), it is possible to obtain the following expression for $\omega_1$:
\begin{eqnarray}
\omega_1 &=& \frac{1}{3}\left[\frac{3b^2\Omega'_D}{1 - \Omega_k - \Omega_D} + \frac{3b^2\Omega_D\left(\Omega'_D+\Omega'_k\right)}{\left(1 - \Omega_k - \Omega_D\right)^2}+\frac{\Omega''_D}{\Omega_D} \right. \nonumber\\
&-& \left. \frac{\Omega'^2_D}{\Omega^2_D}+\frac{\Omega''_D+\Omega''_k}{1 - \Omega_k - \Omega_D}+\frac{\left(\Omega'_D+\Omega'_k\right)^2}{\left(1 - \Omega_k - \Omega_D\right)^2}\right]_0. \label{83}
\end{eqnarray}
We now want to find the explicit forms of the parameters $\omega_0$ and $\omega_1$.\\
Adding Eqs. (\ref{48}) and (\ref{75}), we have:
\begin{eqnarray}
\Omega_D' + \Omega_k' &=& \left( 1-\Omega_D - \Omega_k  \right) +2\left( 1-\Omega_D  -\Omega_k \right)\left( 1 + \frac{1}{H}  \right) -3b^2 \Omega_D = \nonumber \\
       &&\left( 1-\Omega_D - \Omega_k  \right)\left( 3 + \frac{2}{H}  \right)-3b^2 \Omega_D. \label{84}
\end{eqnarray}
Dividing Eq. (\ref{75}) by the fractional energy density of DE $\Omega_D$, we have:
\begin{eqnarray}
 \frac{\Omega_D'}{\Omega_D}= \frac{\left( 1-\Omega_D - \Omega_k  \right)}{\Omega_D} +2\left( \frac{1-\Omega_D}{\Omega_D}   \right)\left( 1 + \frac{1}{H}  \right) -3b^2. \label{85}
\end{eqnarray}
Differentiating Eqs. (\ref{48}) and (\ref{75}) with respect to $x = \ln a$, we obtain the following expressions for $\Omega_D''$ and $ \Omega_k''$:
\begin{eqnarray}
 \Omega_D''&=&- \left( \Omega_D' + \Omega_k'  \right)-2\Omega_D' \left( 1+ \frac{1}{H}  \right) - \frac{2}{H^2}\left( 1-\Omega_D\right)-3b^2\Omega_D', \label{86}\\
\Omega_k''&=& -2\Omega_k'\left( 1+\frac{1}{H}  \right) + \frac{2\Omega_k}{H^2}. \label{87}
\end{eqnarray}
Adding Eqs. (\ref{86}) and (\ref{87}), we obtain:
\begin{eqnarray}
\Omega_D'' +  \Omega_k''  &=& -\left(\Omega_D'  +\Omega'_k  \right)\left( 3+ \frac{2}{H}  \right) - \frac{2}{H^2}\left( 1-\Omega_D -\Omega_k\right)-3b^2\Omega_D' . \label{88}
\end{eqnarray}
Inserting Eqs. (\ref{84}) and (\ref{85}) into the expression of $\omega_0$ given in Eq. (\ref{82}), we obtain the same expression of $\omega_0$ as in the non interacting case.\\
Moreover, using Eqs. (\ref{75}), (\ref{84}), (\ref{85}), (\ref{86}),  (\ref{87})  and (\ref{88}) in Eq. (\ref{83}), we obtain the following expression for $\omega_1$:
\begin{eqnarray}
\omega_1 &=& \frac{1}{3}\left \{    \frac{3b^2    \left[\left( 1-\Omega_D - \Omega_k  \right) +2\left( 1-\Omega_D   \right)\left( 1 + \frac{1}{H}  \right) -3b^2 \Omega_D \right] }{1 - \Omega_k - \Omega_D} +    \right. \nonumber\\
&& \left.  \frac{3b^2\Omega_D}{\left(1 - \Omega_k - \Omega_D\right)^2}\times   \left[\left( 1-\Omega_D - \Omega_k  \right)\left( 3 + \frac{2}{H}  \right)-3b^2 \Omega_D\right] +  \right. \nonumber\\
 &&\left.  \frac{- \left( \Omega_D' + \Omega_k'  \right)-2\Omega_D' \left( 1+ \frac{1}{H}  \right) - \frac{2}{H^2}\left( 1-\Omega_D\right)-3b^2\Omega_D'}{\Omega_D} -  \right. \nonumber\\
 &&\left. \frac{\left[\left( 1-\Omega_D - \Omega_k  \right)\left( 3 + \frac{2}{H}  \right)-3b^2 \Omega_D\right]^2}{\Omega^2_D}- \right. \nonumber\\
 &&\left. \frac{\left(\Omega_D'  +\Omega'_k  \right)\left( 3+ \frac{2}{H}  \right) + \frac{2}{H^2}\left( 1-\Omega_D -\Omega_k\right)+3b^2\Omega_D' }{1 - \Omega_k - \Omega_D}+ \right. \nonumber\\
 &&\left. \frac{\left[\left( 1-\Omega_D - \Omega_k  \right)\left( 3 + \frac{2}{H}  \right)-3b^2 \Omega_D\right]^2}{\left(1 - \Omega_k - \Omega_D\right)^2}\right \}_0. \label{83final}
\end{eqnarray}
For completeness, we now derive the expression of the deceleration parameter $q$, which is generally defined as follow:
\begin{eqnarray}
q=-\frac{\ddot{a}a}{\dot{a}^2}=  -\frac{\ddot{a}}{aH^2}  =-1-\frac{\dot{H}}{H^2}. \label{89}
\end{eqnarray}
The deceleration parameter $q$ can be used in order to quantify the status of the acceleration of the universe \cite{dabro}. The expansion of the universe results to be accelerating if $\ddot{a}$ is positive (as recent cosmological measurements suggest), and in this case $q$ assumes a negative value, whereas positive values of the present day value of $q$ indicates a universe which is either decelerating or expanding at the coasting  \cite{alam}. The minus sign and the definition of deceleration parameter have an historical basis.\\
Dividing the Friedmann given in Eq. (\ref{27}) by $H^2$ and using Eqs. (\ref{28}), (\ref{29}) and (\ref{30}), it is possible to write the deceleration parameter as follow:
\begin{eqnarray}
q=\frac{1}{2}\left[1 - \Omega_k + 3\Omega_D \omega_D  \right].\label{90}
\end{eqnarray}
In order to find the final expressions of $q$ for the RDE models we are studying, we just need to insert in Eq. (\ref{90}) the expressions of the EoS parameter obtained in Eq. (\ref{37}) with the relevant expression of $\rho_D$ corresponding to the particular model.\\
We can make some considerations about the present day value of the deceleration parameter, i.e. $q_0$.
Since we are considering the present day universe, we can consider the power law and the logarithmic corrections practically negligible and consider only the first term in the energy densities given in Eqs. (\ref{7}) and (\ref{8})
Using the expression of $G_{cosmo}$ given in Eq. (\ref{23}) along with the values
of $\Omega_{k0}$, $\Omega_{D0}$ and the value of $\alpha$ obtained in the work of Gao et al., i.e. 0,46, we derive the following relation between $q_0$ and $\lambda$:
\begin{eqnarray}
q_0 = 1.7 - \frac{\lambda}{0.46}.
\end{eqnarray}
From the above equation we easily derive that, for $\lambda > 0.782$, the expression of $q_0$ leads to an
accelerating universe.

\section{Conclusions}
In this paper, we considered the power-law and the logarithmic entropy corrected versions of the RDE model and we have investigated them in a FLRW universe in the framework of Ho\v{r}ava-Lifshitz gravity for both non-interacting and interacting dark sectors.\\
We calculated the general expression of the equation of state (EoS) parameter $\omega_D$ then, using a low redshift expansion of the EoS parameter of DE as $\omega_D \left( z \right) = \omega_0 + \omega_1 z$, we calculated, for the case corresponding to absence and later on presence of interaction between Dark Sectors, both the expressions of $\omega_0$ and $\omega_1$ as functions of the DE and curvature density parameters (and of the interaction parameter $b^2$ in presence of interaction).
We interestingly found the same equation for $\omega_0$ in both non-interacting and interacting DE and DM, which implies that in the interacting case $\omega_0$ has not dependence from the interaction parameter $b^2$.
Instead, the parameter $\omega_1$, in the case of interacting dark sectors, has a clear dependence from the interaction parameter  $b^2$.\\
We also calculated the expression of the deceleration parameter $q$ as function of the equation od state (EoS) parameter $\omega_D$. Moreover, studying the present day value of $q$, i.e. $q_0$, we found that it leads to an accelerating universe for values of the parameter $\lambda$ of the HL cosmology greater than 0.782.

\end{document}